\documentclass[conference]{IEEEtran}
\IEEEoverridecommandlockouts
\usepackage{cite}
\usepackage{amsmath,amssymb,amsfonts}
\usepackage{algorithmic}
\usepackage[linesnumbered,ruled]{algorithm2e}
\usepackage{graphicx}
\usepackage{textcomp}
\usepackage{xcolor}
\usepackage{multirow}
\usepackage{xcolor,colortbl}
\definecolor{green}{rgb}{0.5,0.9,0.0}

\def\BibTeX{{\rm B\kern-.05em{\sc i\kern-.025em b}\kern-.08em
    T\kern-.1667em\lower.7ex\hbox{E}\kern-.125emX}}

\renewcommand\baselinestretch{0.8698}
\begin{document}
\date{}

\setlength{\abovedisplayskip}{2pt}
\setlength{\belowdisplayskip}{2pt}
\setlength{\textfloatsep}{3pt plus 1.0pt minus 1.0pt}
\setlength{\floatsep}{3pt plus 1.0pt minus 1.0pt}

\title{\Large\textbf{Energy-aware Scheduling of Jobs in Heterogeneous Cluster Systems Using Deep Reinforcement Learning}\vspace{-2mm}}	


\author{\IEEEauthorblockN{Amirhossein Esmaili, Massoud Pedram}
\IEEEauthorblockA{Department of Electrical and Computer Engineering,
University of Southern California\\
Los Angeles, California\\
Email: esmailid@usc.edu,
pedram@usc.edu}\vspace{-10mm}}

\maketitle
\thispagestyle{empty}

{\small\textbf{Abstract---
Energy consumption is one of the most critical concerns in designing computing devices, ranging from portable embedded systems to computer cluster systems. Furthermore, in the past decade, cluster systems have increasingly risen as popular platforms to run computing-intensive real-time applications in which the performance is of great importance. However, due to different characteristics of real-time workloads, developing general job scheduling solutions that efficiently address both energy consumption and performance in real-time cluster systems is a challenging problem. In this paper, inspired by recent advances in applying deep reinforcement learning for resource management problems, we present the Deep-EAS scheduler that learns efficient energy-aware scheduling strategies for workloads with different characteristics without initially knowing anything about the scheduling task at hand. Results show that Deep-EAS converges quickly, and performs better compared to standard manually-tuned heuristics, especially in heavy load conditions.
}}

\section{Introduction}
Energy efficiency in cluster systems is an important design factor, as it not only can reduce the operational electricity cost, but also can increase system reliability. Furthermore, these platforms are becoming more popular for many computing-intensive real-time applications such as image or signal processing, weather forecasting, and so forth \cite{b1,greek,b2}. A major portion of this trend is due to rapid progress in computing power of commodity hardware components and their relatively low cost \cite{b2}. Therefore, developing scheduling strategies that achieve promising performance metrics for real-time workloads while yielding low energy costs are of great necessity.

Traditionally, majority of these scheduling problems are solved today using carefully designed heuristics, as they are usually combinatorial NP-hard problems \cite{b4}. There are several works in the literature addressing energy-aware scheduling problem for heterogeneous clusters \cite{greek, b2, Li, zhang}. Authors in \cite{zhang} propose energy-aware task scheduling solutions on DVS-enabled heterogeneous clusters based on an iterated local search method (DVS: dynamic voltage scaling). In \cite{b2}, authors present an adaptive energy-aware scheduling of jobs on heterogeneous clusters with the goal of making the best trade-offs between energy conservation and admissions of subsequently arriving tasks. Generally, the main approach in these studies is developing clever heuristics that have performance  guarantee under certain conditions, which in some cases is followed by further testing and tuning for  obtaining a better performance in practice.

Inspired by recent advances in employing reinforcement learning (RL) for addressing resource management problems, in this paper, we examine building intelligent systems which learn by their own to achieve energy-aware scheduling strategies, as an alternative to using manually-tuned heuristics. While major portion of successful machine learning techniques fall into the category of supervised learning, in which a mapping from training inputs to outputs is learned, supervised learning cannot be applicable to most combinatorial optimization problems, such as nontrivial scheduling problems, as optimal labels are not available due to inherent NP-hardness of most of these problems in nontrivial settings. However, one can evaluate the performance of a set of solutions using a verifier, and provide some feedbacks to a learning algorithm. Consequently, approaching a combinatorial optimization problem using an RL paradigm, could be promising \cite{b9}.

\begin{figure}
\centerline{\includegraphics[width=0.45\textwidth]{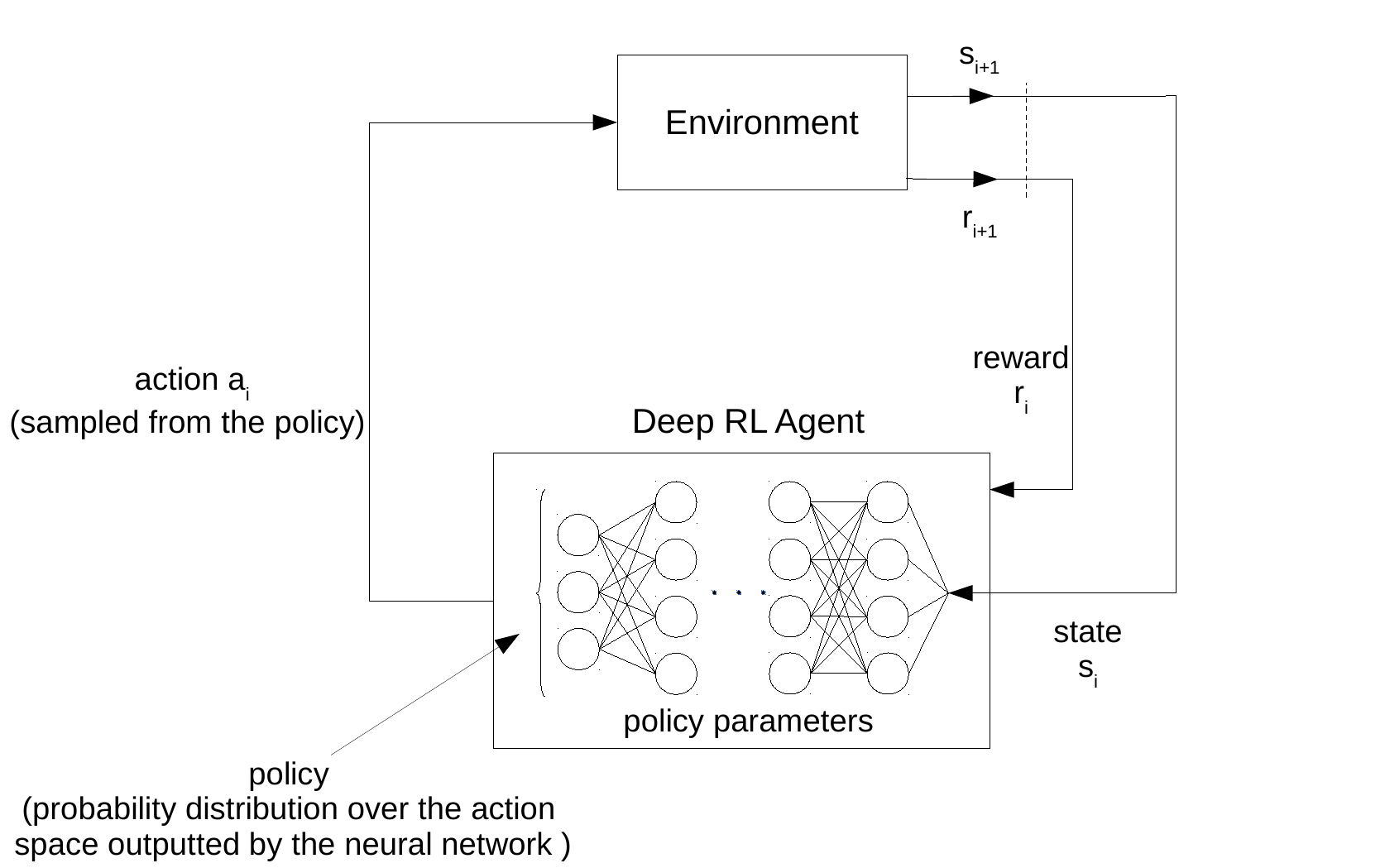}}
\caption{\small A high-level view of the reinforcement learning with the policy represented by a deep neural network. 
}
\label{DRL_high_level}
\end{figure}

In general, RL agent start from not knowing anything from the task at hand, and improves itself based on how well it is doing in the system. Particularly, we approach the problem with the help of deep RL. A high-level view of how deep RL works is shown in Fig. \ref{DRL_high_level}. In each step $i$, the deep RL agent observes a state $s_i$, and performs an action $a_i$. This action is sampled from a probability distribution over the action space, where this distribution is obtained by the underlying neural network with parameters $\theta$ given the state $s_i$ as its input, and is referred to as the \textit{policy} of the deep RL agent shown by $\pi_\theta(s,a)$, where $\pi_\theta(s,a)$ is the probability that action $a$ is taken in state $s$. Therefore, $\pi_\theta(s,a) \rightarrow [0,1]$. $\theta$ is referred to as the \textit{policy parameters} of the agent. Following the action $a_i$, the system state would change to $s_{i+1}$ and a reward $r_{i+1}$ is given to agent. The agent has only control on what action it can do, and not on the obtained reward or state transition. During training, by performing a series of interactions with the environment, the parameters of the underlying neural network will be adjusted for the goal of improving the policy and maximizing the expected cumulative discounted reward: $\mathrm{E}[\sum_{i=0}^{\infty}\gamma^i{r_i}]$, in which $\gamma \in (0,1]$ is the discount factor representing how much the agent cares about the future rewards. RL has been recently combined with deep neural networks to be effective in applications with large space of state and action pairs. In those applications, storing the policy in tabular form would not be feasible anymore and function approximators with tunable parameters, such as deep neural networks, are commonly used \cite{b5, b7, b8}.

The main motivation for the proposed method compared to prior work in energy-aware scheduling for heterogeneous clusters is that the proposed Deep-EAS agent starts from knowing nothing about the scheduling task at hand, and learns nontrivial scheduling policies by modeling the different aspects of the system such as the arrival rate, duration and resource-demand profile of incoming jobs, current occupation state of servers and energy profile of using each one for scheduling any of the waiting jobs, and so forth. The obtained scheduling strategy can be employed in an online scheduling environment and be efficient under varying workload conditions as we see in Section \ref{results}.

The proposed method in this paper uses the notions similar to ones used in \cite{b4}, which is the first successful attempt to our knowledge that solely using deep RL, addresses the conventional problem of scheduling for multi-resource constrained jobs in clusters. However, \cite{b4} does not consider heterogeneity of computing machines in terms of their energy profile in the cluster and thus does not examine energy awareness in its proposed scheduling solution. There are some challenges associated with crafting the rewards function in RL formulation so that the scheduling solution would be energy-aware, which are explained in detail in Section \ref{RL}. Furthermore, in \cite{b4}, it is assumed that the duration of incoming jobs is known upon arrival. However, in a realistic scenario, uncertainties can occur due to miss-predictions on the workloads \cite{Li}. Therefore, the proposed method also takes into account the uncertainties associated with the workloads of arriving jobs.

Consequently, in this paper, using the deep RL paradigm, we present \textit{Deep-EAS}, an online energy-aware scheduler for cluster systems that have multiple machines with heterogeneous energy profiles. The detailed model of the underlying cluster system and associated RL formulation will be presented in Section \ref{model} and Section \ref{RL}, respectively. In Section \ref{background}, a detailed explanation on how Deep-EAS is trained will be presented. In Section \ref{results}, we compare Deep-EAS with comparable heuristics under varying workload conditions and examine the situations where using Deep-EAS is advantageous compared to manual heuristics. Finally, Section \ref{conclusion} concludes the paper.

\section{Method} \label{sec_method}
\subsection{Cluster Model and the Objective Function} \label{sec_cluster}
We consider a cluster with $K$ heterogeneous machines in terms of different energy profiles. Each machine is comprised of $N$ processors that can serve the jobs requiring multiple processors for their execution. Jobs arrive to the system in an online fashion in discrete timesteps. Energy profile of machine $k$ for job $j$ is shown by $e_{j,k}$, which represents the normalized energy consumption of one processor in machine $k$ for job $j$ in one timestep, if that processor is invoked for execution of the job in that timestep.
The number of processors required for the execution of job $j$ is represented by $n_j$.
For determining the actual duration of job $j$ on machine $k$, represented with $d_{j,k}$, similar to \cite{Li}, we assume that the duration profile is known only in advance as probability distribution such as normal distribution, i.e., $d_{j,k} \sim \mathcal{N}(\mu_{j,k},\,\sigma_{j,k}^{2})\,$, where $\mu_{j,k}$ and $\sigma_{j,k}^{2}$ represent the expected value and variance of $d_{j,k}$, respectively. We assume $\sigma_{j,k}^{2}$ to be a ratio of $\mu_{j,k}$, i.e., $\sigma_{j,k}^{2} = \frac{\mu_{j,k}}{c}$, where coefficient $c$ reflects the accuracy of workload estimator of incoming jobs.  

For each job $j$, $\mu_{j,k}$ on machines with higher performance (operating frequency), and correspondingly higher energy profile, is lower than the machines with lower energy profiles. For instance, if for a job $j$ and two machines in the cluster such as machine 0 and machine 1, we have $e_{j,0} > e_{j,1}$, then we will have $\mu_{j,0} < \mu_{j,1}$.

The scheduler, in each discrete timestep, selects and assigns a number of jobs to machines from a queue of waiting jobs. $\Pi_j$ represents the machine that job $j$ is assigned to by the Deep-EAS agent ($0 \leq \Pi_j < K$). A job $j$ assigned to machine $k$ is executed until the end of $d_{j,k}$.
Furthermore, $n_j$ processors are allocated continuously for the entire execution span of job $j$. As we will see in Section \ref{section_efficiency}, even with these assumptions, the Deep-EAS agent provides nontrivial solutions that are advantageous compared to manually-tuned heuristics, especially in heavy load conditions.

As both energy and performance should be addressed, we aim for optimizing the average normalized energy-delay product for arriving jobs. The normalized delay for job $j$ is represented by $D^{norm}_{j}$, and is calculated as follows: $D^{norm}_{j}=\frac{D_j}{\mu_{j,*}}$, in which $D_j$ represents the time it takes from the arrival of job $j$ until its execution completion and departure from the system (including the waiting time of the job in the waiting queue), and $\mu_{j,*}$ represents the minimum $\mu_{j,k}$ among all machines. 
Normalizing $D_j$ prevents biasing the solution towards longer jobs. The normalized energy consumption associated with the complete execution of job $j$ is represented by
\begin{equation} \label{eq.ej}
E_{j,{\Pi_j}}=n_j \times e_{j,{\Pi_j}} \times \frac{d_{j,{\Pi_j}}}{{\mu_{j,*}}}.    
\end{equation}
 Therefore, our scheduling goal is minimizing $\mathrm{E}[E_j \times D_{j}^{norm}]$, where the expectation is calculated over all jobs in the job arrival sequence.

\label{model}

\setlength{\belowcaptionskip}{-28pt}
\begin{figure*}
  \begin{center}
  \includegraphics[width=0.9\textwidth]{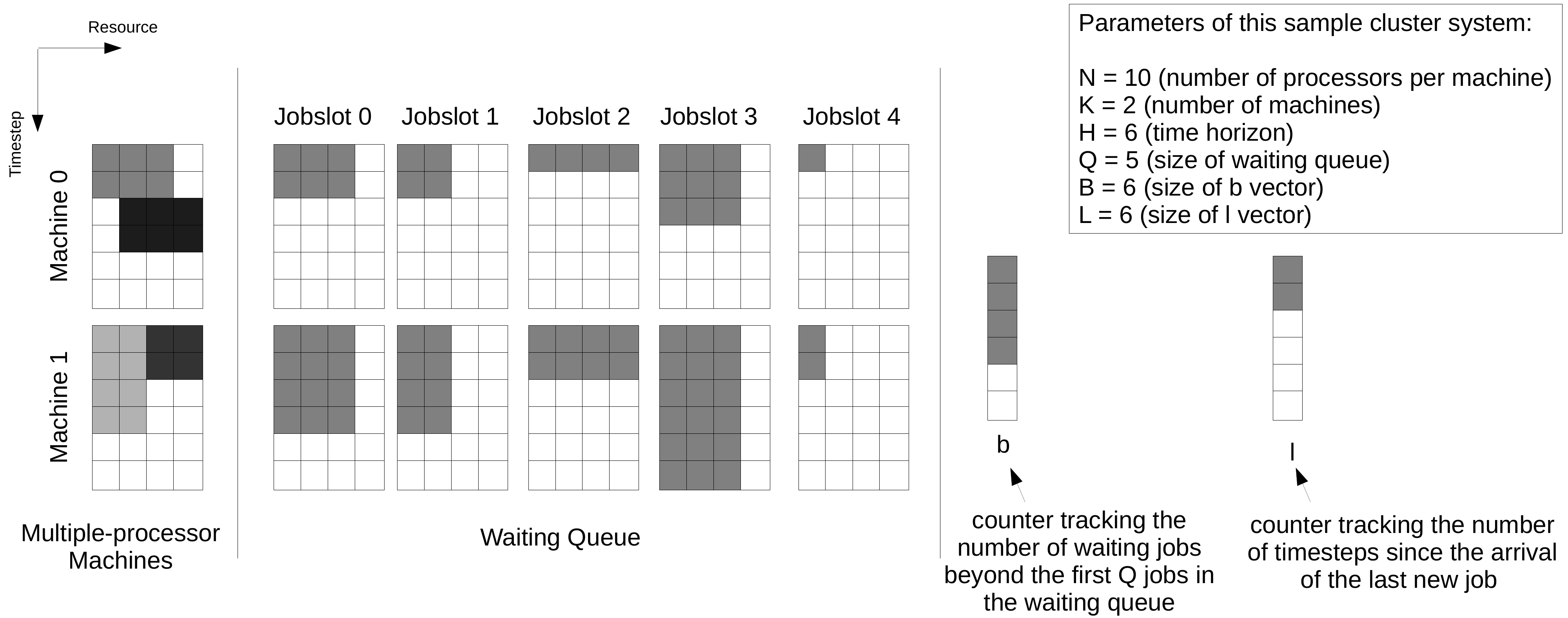}
  \caption{\small An illustrative example of the state representation of the cluster system in the middle of the job arrival process.
  }
  \label{state_rep}
  \end{center}
  \vspace{-6mm}
\end{figure*}
\setlength{\belowcaptionskip}{3pt}

\subsection{Deep RL Formulation for Deep-EAS Agent} \label{RL}
\subsubsection{State Space}
For state representation of the system in each timestep, we represent the current occupation state of machines and the resource-demand and average duration profile of the jobs in the waiting queue as \textit{binary matrices}. Fig. \ref{state_rep} illustrates a sample state of a system with two machines and a waiting queue of size five. The matrices corresponding to the machines, shown on the left side of Fig. \ref{state_rep}, represent the occupation state of these machines from the current timestep until $H$ timesteps ahead in the future. For instance in Fig. \ref{state_rep}, two jobs are scheduled on machine 1. For the sake of argument, we refer to these jobs as job 0 and job 1. Job 0 uses two processors for the next $d_{0,1}$ timesteps where $d_{0,1} \sim \mathcal{N}(4,\,4/c)\,$, and job 1 uses two processors for the next $d_{1,1}$ timesteps where $d_{1,1} \sim \mathcal{N}(2,\,2/c)\,$.

Furthermore, the matrices corresponding to the jobs in the waiting queue, shown in the middle of Fig. \ref{state_rep}, represents the resource-demand and average duration profile of the jobs in the waiting queue on each of machines. 
The average duration profile of each job on different machines are different. For instance, in Fig. \ref{state_rep} that we have two heterogeneous machines, we see two different duration profiles for each job in the waiting queue. In this case, the duration profile of each job on the machine 1 is higher compared to the duration profile of that job on the machine 0, which conveys the fact that in this sample system, machine 1 is the machine with the lower performance (operating frequency) and lower energy profile compared to machine 0. The  size  of  the  queue  is represented by $Q$.

In order to have a finite state representation, we maintain the binary matrices corresponding to the resource-demand and average duration profile of waiting jobs only for the first $Q$ jobs arrived at the system that have not yet been scheduled on any of the machines. For the further jobs, we incorporate only their count in the state of the system. We use a binary vector $b$ for representing these backlog jobs, in which the number of 1s in this vector represents the count of backlog jobs. Furthermore, for making the scheduler aware of the arrival rate of incoming jobs, we track the number of discrete timesteps since the last job has arrived to the system. We use a binary vector $l$ in which the number of 1s represents the count of timesteps since the arrival of last new job. The length of vectors $b$ and $l$, shown on the right side of Fig. \ref{state_rep}, are represented by $B$ and $L$, respectively, and should be large enough so that these vectors do not get exhausted. $B$ and $L$ are also chosen to be integer multiples of time horizon $H$, as we want to tile the vectors $b$ and $l$ in chunks of size $H$ so that we have a rectangular state matrix.
Consequently, the height and width of the state binary matrix, that incorporates all the mentioned information, would be obtained as $H$ and $N \times (1+Q) \times K + \frac{B}{H} + \frac{L}{H}$, respectively. 
\subsubsection{Action Space}
In each timestep, the scheduler can potentially select one or more jobs from the waiting queue with the size of $Q$ and assigns it to one of $K$ machines. Therefore, the size of the action space would become exponentially large with respect to $Q$ and $K$. In order to reduce the action space size,
similar to \cite{b4}, we decouple the decision steps of the Deep-EAS agent from real timesteps, and allow the agent to do multiple actions in a single timestep. The new action space is associated with selecting one of the waiting jobs and assigning it to one of $K$ machines. Therefore, the size of the action space is reduced to $Q \times K$. Specifically, we define the action $k \times Q+q$ as ``assign the job in $q$-th slot in the waiting queue to machine $k$", where $0 \leq q < Q$ and $0 \leq k < K$. We define the action $K \times Q$ as the ``hold" action. Upon taking this action, the agent does not schedule any further jobs in the current timestep (by considering hold action, the actual action space size would be $Q \times K+1$ instead of $Q \times K$). In each timestep, the scheduler can take multiple actions until choosing the hold action, or an \textit{invalid action}. The selected action is invalid if there is no job in $q$-th slot of the waiting queue, or if the selected job does not fit in the selected machine from the current timestep looking ahead into the next $H$ timesteps based on the average estimates available for jobs occupying the underlying machines. 

By choosing each valid action, the corresponding job in the queue is assigned to the selected machine starting from the earliest possible timestep on that machine, and a job from the backlog queue (if any) is dequeued and replaces the job that has just been scheduled. However, by choosing an invalid action or the hold action, the time actually goes on and the state matrix shifts up by one row and jobs which have finished their execution, according to their actual duration sampled from the corresponding normal distribution, depart from the system. Therefore, jobs may depart from the system sooner or later than their average estimates. This will either advance or postpone the actual start time of other jobs waiting for resources to get freed up. Furthermore, when the actual time proceeds, any new jobs may arrive to the system, depending on the job arrival process. If any new job arrives, vector $l$ resets to an all-zero vector.
\subsubsection{Rewards Function} \label{sec_rewards}
 One challenge for defining the rewards function for our problem is the fact that for the jobs in the waiting queue and backlog, in the timesteps before they actually get assigned to one of the underlying machines, we know their contribution to the average delay of jobs (one for every timestep they are still in the system). However, we do not know their corresponding $E_j$ before they get assigned to one of the machines. 
In order for the cumulative rewards function to correlate with our objective, normalized energy-delay product, we need to weight each timestep that each job $j$ is still in the system with 
$\frac{E_j}{\mu_{j,*}}$ (see Section \ref{sec_cluster}). For solving this issue, we define the rewards function in each timestep as the following (no reward is given for intermediate actions of the scheduler agent during a timestep. Reward is only granted after the actual time proceeds):
\begin{equation} \label{eq_reward}
-\left(\sum_{j \in J_p}{\frac{E_{j}}{\mu_{j,*}}} + \sum_{j \notin J_p}{\frac{E^{*}_{j}}{\mu_{j,*}}} + \sum_{j \in J_{new}}{\frac{\delta^{correct}_j}{\mu_{j,*}}}\right).
\end{equation}
The breakdown of three terms of the \eqref{eq_reward} are as follows:

    \textbf{First term}: 
    $J_p$ represents the set of jobs currently scheduled on any of machines. For each job $j$ in this set, we know the energy consumption associated with execution of job $j$, $E_j$. 
    
    \textbf{Second term}: 
    For each job $j$ which is not currently scheduled on any of machines, we do not know yet their energy consumption. For such job $j$, we temporarily assume we will eventually assign it to the machine that yield the minimum energy consumption for its execution, and represent this value by $E^{*}_{j}$. We will correct our assumption using the to-be-explained third component of the rewards function, when we eventually assign the job to one of the underlying machines. 
    
    \textbf{Third term}: 
    $J_{new}$ represents the set of jobs that have been ``just" scheduled on a machine in the current timestep. For each job $j$ in $J_{new}$, we have used the second term of \eqref{eq_reward} during previous timesteps from the time the job arrived to the system. In case the current assigned machine of job $j$ is not the machine that yields the lowest energy consumption for job $j$, which was our temporary assumption in the second component of \eqref{eq_reward}, we correct and add the amount of difference for previous timesteps to the rewards function. This amount for such job $j$ is represented by $\delta^{correct}_j=(E_{j} - E^{*}_{j}) \times |\Delta t|$, where $|\Delta t|$ represents the number of timesteps from the time job arrived to the system until the current timestep ($|\Delta t|$ represents just the number of timesteps and is unit-less itself). 
    
    Consequently, using the discount factor $\gamma = 1$, the cumulative rewards function \eqref{eq_reward} over all timesteps would result the (negative) total of normalized energy-delay product over all the jobs, and maximizing this cumulative reward results in minimizing the total and thus the average of normalized energy-delay product over all the jobs.

\subsection{Training Deep-EAS} \label{background}
For training the Deep-EAS agent, we need to adjust the policy parameters of its underlying deep neural network (see Fig. \ref{DRL_high_level}). Similar to \cite{b5}, we use \textit{policy gradients} in which we learn by employing gradient descent on the policy parameters. For using gradient descent, we need to have the gradient of the expected cumulative discounted reward, $\mathrm{E}[\sum_{i=0}^{\infty}\gamma^i{r_i}]$, which is our objective function. This gradient is obtained using the REINFORCE equation \cite{williams}:
\begin{equation} \label{eq_gradient}
\nabla_\theta\mathrm{E}_{\pi_\theta}[\sum_{i=0}^{\infty}\gamma^i{r_i}] = \mathrm{E}_{\pi_\theta}[R_{\pi_\theta}(s,a).\nabla_\theta\log\pi_\theta(s,a)],
\end{equation}
where $R_{\pi_\theta}(s,a)$ represents the expected cumulative discounted reward if we choose action $a$ in state $s$ and follow the policy $\pi_\theta$ afterwards. In policy gradients, in each training iteration, the main idea is that we approximate the gradient equation in \eqref{eq_gradient} by evaluating the trajectories of executions obtained by following the policy we have in that iteration. Specifically, for training Deep-EAS using policy gradients, in each training iteration, we draw a number of trajectories of the executions sampled from $\pi_\theta$ for a sample job arrival sequence. Each execution trajectory (episode) terminates when all the jobs in the sequence finish their execution (or a predefined maximum length of the trajectory is reached). To train a generalized policy, we use multiple sample job arrival sequences in each training iteration ($S$ sequences), and we perform $M$ trajectories of execution for each sequence until the trajectory termination. Using these trajectories, we approximate \eqref{eq_gradient} as follows:
\begin{equation} \label{eq_gradient_approx}
\begin{split}
      &\nabla_\theta\mathrm{E}_{\pi_\theta}[\sum_{i=0}^{\infty}\gamma^i{r_i}] \approx \\
      &\frac{1}{S.M}\sum_{s=1}^{S}\sum_{m=1}^{M}{\sum_{t}\nabla_\theta\log\pi_\theta(s^{s,m}_t,a^{s,m}_t)v^{s,m}_t,} 
\end{split}
\end{equation}
in which $v^{s,m}_t$ is the empirically computed cumulative discounted reward and serves as an unbiased estimate of $R_{\pi_\theta}(s^{s,m}_t,a^{s,m}_t)$ (superscript $s$ and $m$ are used to refer to $m$-th trajectory of $s$-th sample job arrival sequence). Using this approximation, we update policy parameters in each iteration via the following equation:
\begin{equation} \label{eq_update}
   \theta \leftarrow \theta + \frac{\alpha}{S.M}\sum_{s=1}^{S}\sum_{m=1}^{M}{\sum_{t}\nabla_\theta\log\pi_\theta(s^{s,m}_t,a^{s,m}_t)(v^{s,m}_t-b^s_t).}     
\end{equation}
$\alpha$ in \eqref{eq_update} indicates the learning rate of the training algorithm. In \eqref{eq_update}, we reduce a baseline value $b^s_t$ from $v^{s,m}_t$ which help reduce the variance of policy gradients. Without reducing the baseline, gradient estimates obtained using \eqref{eq_gradient_approx} can have high variances \cite{Schulman}. For calculating $b^{s}_t$, the average of $v^{s,m}_t$ at the same timestep $t$ over all trajectories ($m=1, 2, ..., M$) of the job sequence $s$ is used.

\section{Evaluation} \label{results}
\subsection{Cluster Setup} \label{cluster_setup}
We use an instance of the cluster system described in Section \ref{sec_method} and shown in Fig. \ref{state_rep} with the following parameters: $K=2$, $N=10$, $H=30t$ ($t$ represents the duration of one timestep), $Q=10$, $B=90$, and $L=30$. Jobs arrive to the system in an online fashion according to a Bernoulli process with the arrival rate of $\lambda$. The length of each job arrival sequence is set to $60t$ (new jobs can arrive until timestep 60, however experiment goes on until all jobs remained in the system beyond $60t$ finish their execution). The resource requirement of each arriving job is chosen uniformly between 1 and 10 processors. In our sample cluster model, we consider machine 0 as the higher-performance machine and machine 1 as the lower-performance machine. Particularly, we consider the operation frequency of machine 0 to be twice the operating frequency of machine 1. Therefore, for each job $j$, we have $\mu_{j,1} = 2\mu_{j,0}$. Furthermore, we consider each job arrival sequence to be a combination of short-duration and long-duration jobs. The probability that an arriving job is a short job is indicated with $\beta$. $\mu_{j,0}$ for each short job $j$ is chosen uniformly between $1t$ and $3t$, while $\mu_{j,0}$ for each long job $j$ is chosen uniformly between $10t$ and $15t$. Coefficient $c$ which was introduced in Section \ref{sec_cluster}, and reflects the accuracy of workload estimator of incoming jobs, is set as 4. We will examine the efficiency of Deep-EAS for different values of $\lambda$, $\beta$, and $c$  in Sections \ref{sec_results} and \ref{sec_beta_c}.

\subsection{Energy Model}
We consider both machines to be always ``on" during the experiment. This means that the energy consumption due to static power consumption of the system serves as an additive factor to the total energy consumption of the system during the experiment. Therefore, the ratio between $e_{j,0}$ and $e_{j,1}$ for each job $j$ needs to reflect the ratio between the dynamic energy consumption of the processor on machine 0 and machine 1 in one timestep.
By employing the power model presented in \cite{amirhossein} and \cite{zhou2017energy}, dynamic power consumption of a processor can be modeled by $x_{j}f^y$, in which $x_j$ is a coefficient depending on the average switched capacitance and the activity factor of job $j$, $f$ is the processor operating frequency, and $y$ is the technology-dependent dynamic power exponent. Therefore, for each job $j$ we have:
$
    \frac{e_{j,0}}{e_{j,1}}= (\frac{f_0}{f_1})^y. 
$
Using a classical energy model of a 70nm technology processor that supports 5 discrete frequencies ranging from $1~GHz$ to $2~GHz$, whose accuracy has been verified by SPICE simulation, \cite{amirhossein} proposes the value for $y$ as 3.2941. Therefore, by setting $f_0=1~GHz$ and $f_1=2~GHz$ (the operating frequencies of our machines) and using this value for $y$, for each job $j$ we have: $\frac{e_{j,0}}{e_{j,1}}=2^{3.2941} =9.809$. While the actual $e_{j,k}$ for the jobs are different with each other due to the different $x_j$ each job $j$ has, the ratio between $e_{j,0}$ and $e_{j,1}$ for each job $j$ remains the same. Therefore, we use the normalized values of $e_{j,0}=9.809$ and $e_{j,1}=1$ for each job $j$. It should be noted that while the energy model based on a 70nm technology is employed here, the proposed method is capable of dealing with a general, parameterized power model. Therefore, in a smaller technology node, one can find the corresponding coefficients and exponents, and use them for finding scheduling solutions.


\subsection{Deep-EAS Training Setup and Overhead Analysis}
For the underlying neural network of the Deep-EAS agent, the size of the input is obtained as a 30 $\times$ 224 binary matrix with the values used for the parameters of the cluster model in Section \ref{cluster_setup}. We apply a convolutional layer to extract features from this matrix. We use eight 3 $\times$ 3 filters with the stride of size 2 (in both height and width directions), followed by the Relu activation function. After this layer, we use a fully connected layer with the size of 21 followed by the softmax activation function (the action space size for our cluster mode is 10 $\times$ 2 + 1). We train Deep-EAS as described in Section \ref{background} using 150 different job arrival sequences for 1000 training iterations. In each training iteration, we evaluate 20 different trajectories of execution for each job sequence. For updating the policy parameters, we use Adam optimizer \cite{adam} with the learning rate of 0.001. 

The deployment of Deep-EAS agent is done after the training is finished. Using our experiment setup, the average latency overhead associated with each inference (scheduling decision) is about $237~\mu S$. This overhead can be considered negligible as the duration of timestep $t$, reflecting the time interval between scheduling decisions, usually takes in the order of a few milliseconds.       

\subsection{Results} \label{sec_results}
As a standard manually-tuned heuristic to compare the proposed Deep-EAS agent with, we choose an energy-aware shortest job first (ESJF) agent. ESJF, in each timestep, schedules the job that yields the lowest normalized energy-delay product to its corresponding machine (according to the available average estimates of duration of jobs in the waiting queue). ESJF keeps doing this process until no job is left in the waiting queue or no further jobs can be scheduled on any of machines in that timestep (due to the occupancy state of machines). In that case, time proceeds and ESJF repeats this procedure. This process continues until all jobs in the job sequence finish their execution.

\begin{figure}
\centerline{\includegraphics[width=0.45\textwidth]{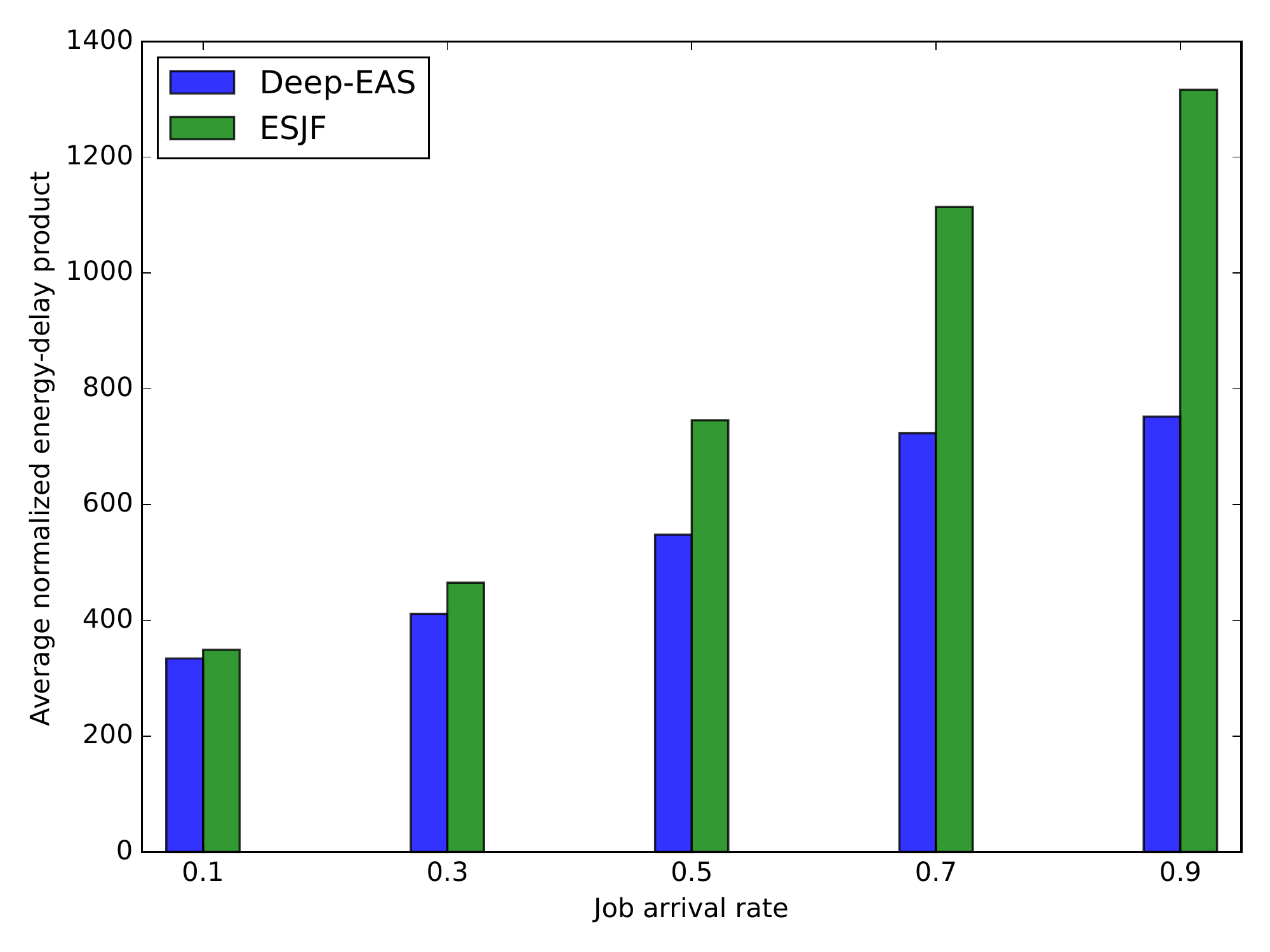}}
\caption{\small Comparison of Deep-EAS and ESJF at different job arrival rates, when $\beta=0.5$.}
\label{diff_job_rate}
\end{figure}

Fig. \ref{diff_job_rate} presents a comparision between Deep-EAS on 150 new jobsets (not seen during training) and ESJF, for different job arrival rates when $\beta=0.5$ (the probability that a new job is a short job is equal to the probability that it is a long job). As presented in Fig. \ref{diff_job_rate}, the average normalized energy-delay product values obtained by either of Deep-EAS and ESJF generally increase with the job arrival rate. Deep-EAS is comparable with ESFJ for low arrival rates (e.g., for $\lambda=0.1$ and $\lambda=0.3$). However, Deep-EAS shows to be considerably advantageous in higher arrival rates. For instance, for $\lambda=0.9$, the average normalized energy-delay product obtained from Deep-EAS is 42.88\% lower in comparison with ESJF.

\subsection{Deep-EAS Training Curve}
\begin{figure}
\centerline{\includegraphics[width=0.45\textwidth]{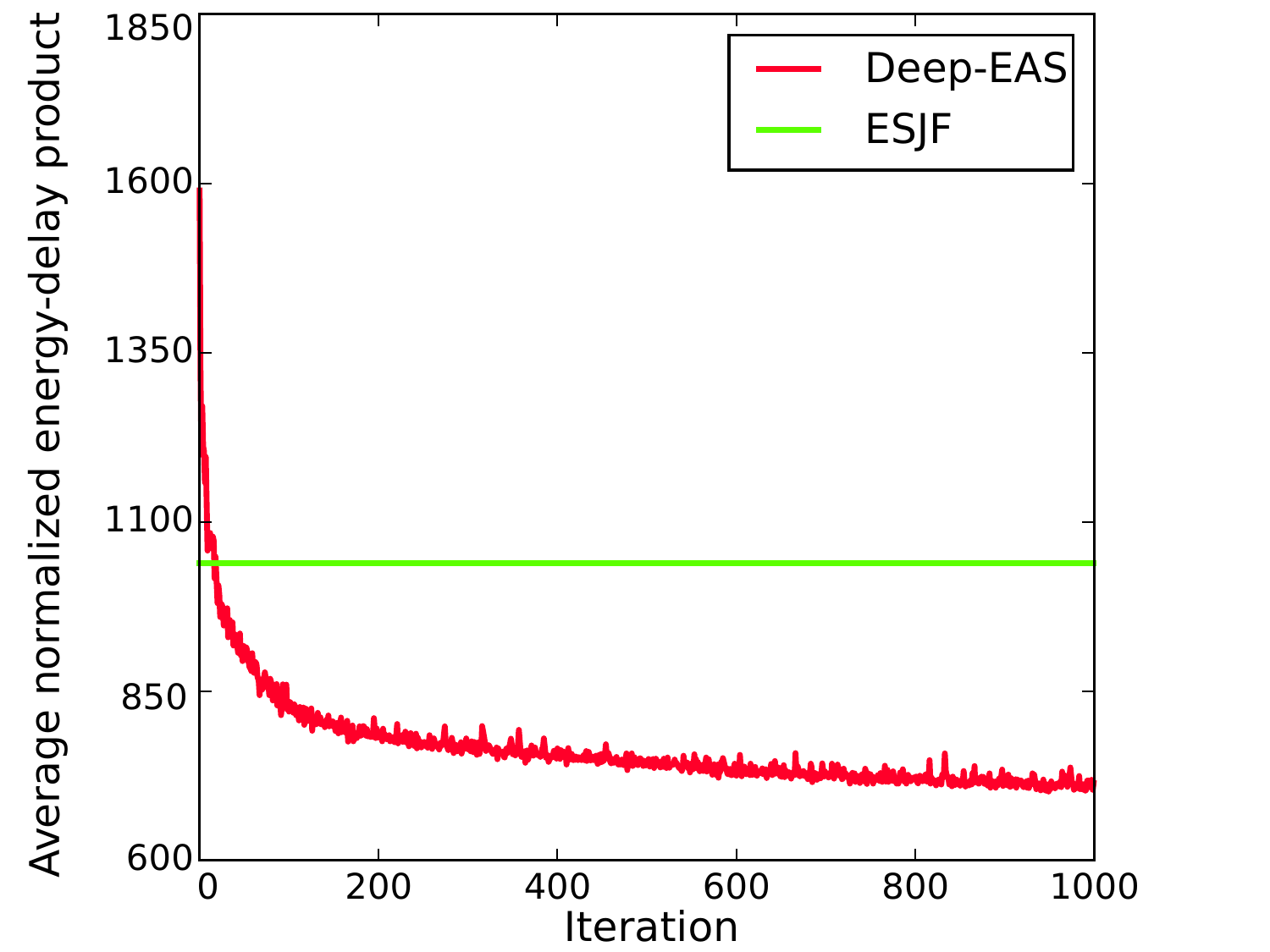}}
\caption{\small Deep-EAS learning curve indicating the policy improvement over the training iterations.}
\label{lr_07_3_n}
\end{figure}

The training curve of Deep-EAS over 1000 iterations and achieved average normalized energy-delay product after each iteration are presented in Fig. \ref{lr_07_3_n}, for the case where the job arrival rate is 0.7 and $\beta=0.5$. The obtained average normalized energy-delay product using ESJF is also shown in Fig. \ref{lr_07_3_n} as a reference. As indicated in Fig. \ref{lr_07_3_n}, Deep-EAS starts from acting poorly in the environment, but quickly improves itself over the training iterations, surpassing the ESJF after the first 30 iterations and further improvement beyond that. For the sake of reducing the time of training, in each training iteration, we performed execution trajectories of each job sequence in parallel on a platform with four 3.2 GHz Intel Core i7-8700 CPUs and 64 GB RAM. On this platform, each training iteration took about 97 seconds on average.

\subsection{Analyzing Why Deep-EAS is Advantageous} \label{section_efficiency}

The main advantage that Deep-EAS possess is that it can develop nontrivial scheduling solutions during training, 
which are \underline{not} necessarily \textit{energy-delay conserving} for every job, or \textit{work conserving} for every timestep. If a scheduling solution is energy-delay conserving for every job, if it allocates a job in a timestep to a machine, it allocates it to the one yielding the minimum normalized energy-delay product for that job. If a scheduling solution is work-conserving for every timestep, it keeps allocating jobs from the waiting queue as long as resources are available in a timestep. ESJF is a scheduling solution which is both energy-delay conserving and work conserving. In general, since manually-tuned resource scheduler solutions usually make decisions in each timestep based on a predefined metric, they are mainly resource conserving \cite{Grandl}. However, Deep-EAS can potentially be both \underline{not} energy-delay conserving and \underline{not} work conserving, if these decisions can eventually cause the lower average normalized energy-delay product over all the jobs. Particularly, for results shown in Fig. \ref{diff_job_rate}, Deep-EAS is not energy-delay conserving for 13.11\% of jobs, and is not work conserving for 91.81\% of timesteps. 

\begin{figure}
\centerline{\includegraphics[width=0.45\textwidth]{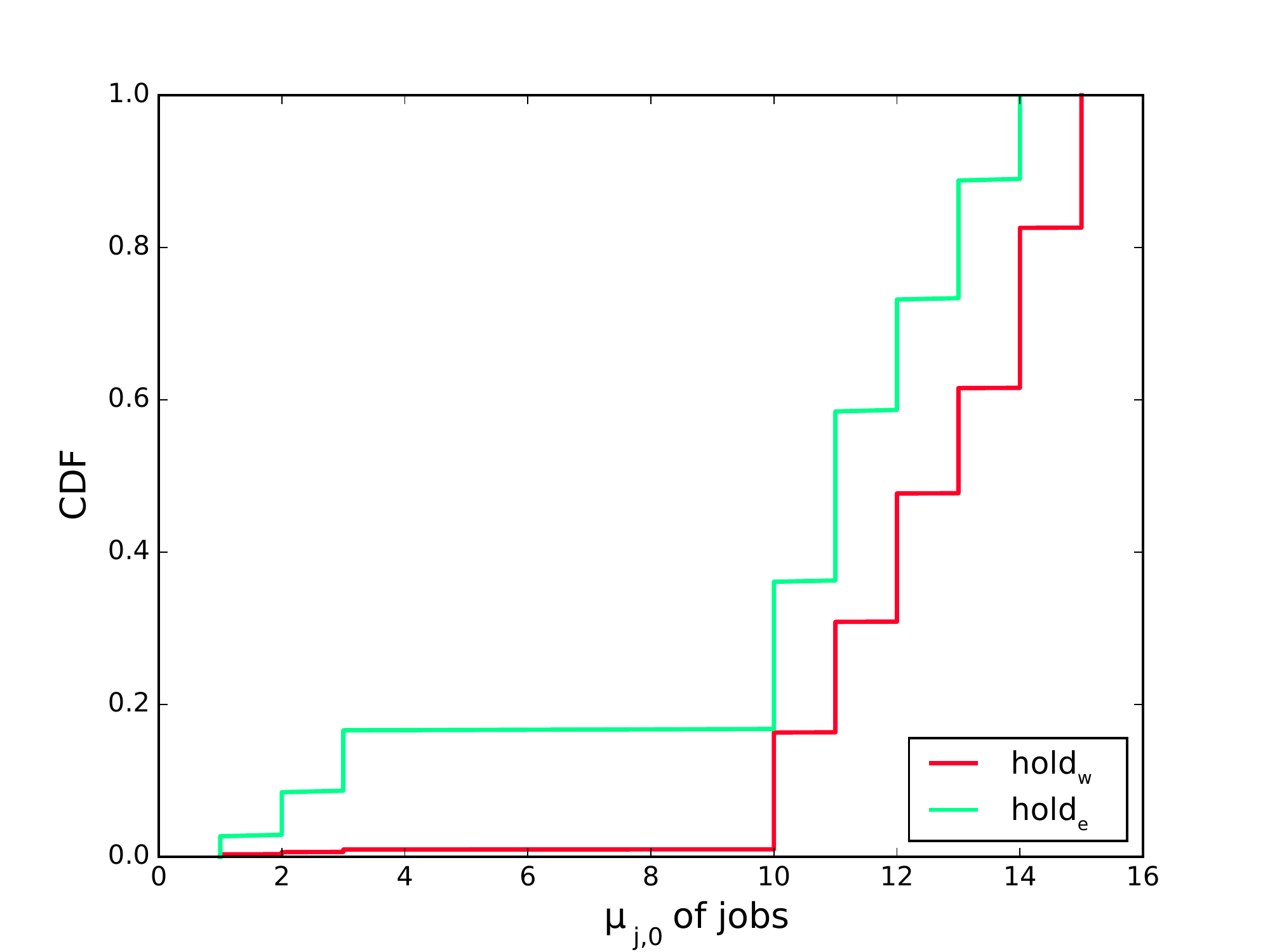}}
\caption{\small CDF plots of $\mu_{j,0}$ of jobs Deep-EAS is not energy-delay conserving for ($hold_e$), alongside the jobs Deep-EAS is not work conserving for ($hold_w$), when $\lambda=0.9$ and $\beta=0.5$.}
\label{cdf_hold}
\end{figure}

To further analyze the scheduled jobs using Deep-EAS, for the case in Fig. \ref{diff_job_rate} where $\beta=0.5$ and $\lambda=0.9$ (a high job arrival rate),
we examine the cumulative distribution function (CDF) plots of the $\mu_{j,0}$ of the jobs Deep-EAS was not energy-delay conserving for (shown with $hold_e$), alongside the jobs Deep-EAS was not work conserving for (shown with $hold_w$).
These CDF plots are shown in Fig. \ref{cdf_hold}. As shown in this figure, while $\beta=0.5$ and thus the number of small jobs and long jobs in a sequence are almost the same, we observe if Deep-EAS does not act as an energy-delay conserving scheduler for a job in a timestep, that job is a long job most of the times (see $hold_e$ in Fig. \ref{cdf_hold}). The intuition behind this could be that allocating long jobs to the machine with lower energy profile (and thus resulting higher duration) could occupy that machine for many timesteps, which can reduce the chance of allocating a number of potentially arriving small jobs to that machine and potentially increase the average normalized energy-delay product over all the jobs. Hence, in a heavy load condition, it can be eventually useful to not be energy-delay conserving for some of the long jobs. Similarly, if Deep-EAS withholds a job in a timestep, that job is most of the times a long job (see $hold_w$ in Fig. \ref{cdf_hold}). Almost the same argument mentioned for $hold_e$ can be presented as the intuition for $hold_w$. In a job arrival process with a high arrival rate, withholding a long job can potentially pave the way for scheduling a number of yet-to-arrive small jobs and eventually being advantageous in reducing the average normalized energy-delay product over \underline{all} the jobs. Deep-EAS learns these solutions on its own.

\subsection{Examining the effect of $\beta$ and $c$} \label{sec_beta_c}
To evaluate the effect of $\beta$, for the case where the job arrival rate is 0.7, we consider 3 cases: $\beta=0.8$ (majority of the jobs are short jobs), $\beta=0.5$ (the number of small jobs and long jobs are almost the same), and $\beta=0.2$ (majority of the jobs are long jobs). By evaluating Deep-EAS and ESJF on 150 new jobsets (not seen during the training of Deep-EAS), for these values of $\beta$, the average normalized energy-delay product obtained by Deep-EAS are 45.29\% ,35.10\%, and 11.94\% lower compared to ESJF. This indicates that for the job sequences where majority of jobs are small jobs, Deep-EAS shows to be more advantageous. 

To evaluate the effect of the workload estimator accuracy, reflected by coefficient $c$ mentioned in Section \ref{sec_cluster}, we reproduce the results in Fig. \ref{diff_job_rate}, but assuming we have a perfect workload estimator of incoming jobs. In other words, for each job $j$ we assume we have $d_{j,k} \sim \mathcal{N}(\mu_{j,k},\,0)\,$ or $d_{j,k}=\mu_{j,k}$. Using this perfect workload estimator, for job arrival rates of 0.1, 0.3, 0.5, 0.7, and 0.9, the obtained average normalized energy-delay product values via Deep-EAS are 5.37\%, 13.84\%, 29.45\%, 37.64\%, and 45.47\% lower, respectively, compared to ESJF. Furthermore, the obtained values for average normalized energy-delay products using the perfect workload estimator showed to be lower for both Deep-EAS and ESJF compared to values in Fig. \ref{diff_job_rate}. Therefore, using a better workload estimator, it is again observed that Deep-EAS shows to be considerably more advantageous in higher arrival rates. However, in all job arrival rates, the gap between values obtained by Deep-EAS and ESJF has been increased.

\setlength{\belowcaptionskip}{3pt}
\setlength{\belowcaptionskip}{3pt}

\section{Conclusions and Future Work} \label{conclusion}
This paper addresses energy-aware scheduling in clusters by proposing Deep-EAS, a scheduler designed with the aid of deep RL. During the training, Deep-EAS starts from knowing nothing about the scheduling task at hand, and develops nontrivial scheduling solutions. We observe these solutions outperform standard manually-tuned heuristics, especially in heavy load conditions with high job arrival rates. For future work, Deep-EAS can be potentially extended to learn more complex strategies such as job-preemption, job-migration and dynamic voltage and frequency scaling (DVFS), which can increase its adaptability to various situations. 

\end{document}